\documentclass[pre,a4,twocolumn,superscriptaddress,showpacs,
preprintnumbers,amsmath,amssymb]{revtex4}
\usepackage{graphicx}
\usepackage{dcolumn}
\usepackage{bm}
\begin{document}
\title{Yukawa particles confined in a channel and subject to a periodic potential: ground state and normal modes}
\author{J.~C.~N.~Carvalho}
\email{joaoclaudio@fisica.ufc.br} \affiliation{Departamento de
F\'isica, Universidade Federal do Cear\'a, Caixa Postal 6030,
Campus do Pici, 60455-760
Fortaleza, Cear\'a, Brazil}%
\author{W.~P.~Ferreira}
\email{wandemberg@fisica.ufc.br} \affiliation{Departamento de
F\'isica, Universidade Federal do Cear\'a, Caixa Postal 6030,
Campus do Pici, 60455-760
Fortaleza, Cear\'a, Brazil}%
\author{G.~A.~Farias}
\email{gil@fisica.ufc.br} \affiliation{Departamento de F\'isica,
Universidade Federal do Cear\'a, Caixa Postal 6030, Campus do
Pici, 60455-760
Fortaleza, Cear\'a, Brazil}%
\author{F.~M.~Peeters}
\email{francois.peeters@ua.ac.be}
 \affiliation{Departamento de F\'isica, Universidade Federal do
Cear\'a, Caixa Postal 6030, Campus do Pici, 60455-760
Fortaleza, Cear\'a, Brazil}%
\affiliation{Department of Physics, University of Antwerp,
Groenenborgerlaan 171, B-2020 Antwerpen, Belgium}
\date{ \today }

\begin{abstract}
We consider a classical system of two-dimensional (2D) charged
particles, which interact through a repulsive Yukawa potential
$exp(-r/\lambda)/r$, confined in a parabolic channel which limits
the motion of the particles in the $y$-direction. Along the
$x$-direction, the particles are also subject to a periodic
potential substrate. The ground state configurations and the
normal mode spectra of the system are obtained as function of the
periodicity and strength of the periodic potential ($V_0$), and
density. An interesting set of tunable ground state configurations
are found, with first and second order structural transitions
between them. A magic configuration with particles aligned in each
minimum of the periodic potential is obtained for
$V_0$ larger than some critical value which has a power law dependence on the density.
The phonon spectrum of different configurations were also calculated. A localization of the
modes into a small frequency interval is observed for a sufficient strength of the periodic
potential. A tunable band-gap is found as a function of $V_0$.
This model system can be viewed as a generalization of the Frenkel
and Kontorova model.
\end{abstract}

\maketitle
\section{ Introduction }
\label{sec:introduction}
Two-dimensional system (2D) are often created in the presence of a
substrate \cite{Grnberg}, which may induce a periodic potential on
the particles. In the pioneering experimental work of Chowdhury
$et$ $al.$ \cite{Chowdhury}, the authors studied a 2D colloidal
system under influence of an one-dimensional (1D) periodic
potential. An optical tweezer was used to trap the colloids by
laser beams. For very high values of light intensity,
crystallization of the colloidal suspension was observed, when the
periodicity of the substrate (periodic potential) was chosen to be
commensurate to the mean particle distance. Laser induced freezing
which is caused by the suppression of thermal fluctuations
transverse to the 1D periodic substrate was found (liquid-solid
transition) \cite{bechinger01}. The system studied in Ref.
\cite{Chowdhury} is related to the colloidal molecular crystal
(CMC) and received attention recently due to important
applications in photonic and phononic crystals
\cite{reich1,reichhardt02}.

Specifically, CMC occurs when the number of colloids is an integer
multiple of the number of substrate minima, and has been investigated using
in simulations \cite{reichhardt02,mikulis04} and realized
experimentally \cite{bechinger02}. CMC is an interesting
experimental system to study order and dynamics in 2D since
typical particle size and relaxation times permit, e.g, to use
digital video-microscopy to track particle trajectories, allowing
a deeper investigation of the physical behavior of the system
\cite{alsayed05}.

Originally, CMC was proposed for 2D system in 2D periodic
potential (substrate). As known the dimensionality of the system
plays an important role in many physical properties of distinct
physical phenomena. In this sense, an interesting question is how the
ordered structures and physical properties would be influenced by
the dimensionality of the periodic substrate. Recently, Herrera-Velarde and Priego
\cite{herrera1,herrera2} studied a 2D system of repulsive
colloidal particles confined in a narrow channel and subject to an
external 1D periodic potential, which could be seen as the 1D
version of the CMC. The main focus of the study was the role of
the substrate in the mechanisms that lead to a variety of
commensurate and non-commensurate phases, its
effect on the the single-file diffusion regime, and the
pinning-depinning transition in 1D systems. The 1D character of
the channel was represented by a hard wall potential. Due to the
repulsive interaction between the particles and the nature of the
confinement potential the density across the channel was found to
be non-uniform with a higher density at the
channel edges.

In the present paper we study the ordered configurations and the
phonon spectrum of a 2D system of repulsive (Yukawa interaction)
particles confined in a parabolic channel and subjected to a 1D
periodic potential along the channel. As compared to the systems
in Refs. \cite{herrera1,herrera2}, and due to the parabolic shape
of the confinement an opposite density distribution is observed,
with particles more concentrated at the central region of the
channel. As shown previously for finite size clusters of repulsive
particles, the confinement potential is determinant, e.g. for
melting and phonon spectra \cite{bedanov94}. The 1D parabolic
confinement introduces a $quasi$-1D (Q1D) character to the system
in the sense that particles are still allowed to move freely in the
perpendicular direction of the confinement potential.

The interplay between the repulsive inter-particle interaction and
the periodic potential determines the different ground state
configurations. Our model system of Yukawa particles can be
realized experimentally using: i) a dusty plasma
\cite{chu1994,liu03,liu05}, ii) colloidal systems
\cite{zahn99,golosovsky02} and iii) electrons on liquid helium
\cite{wigner1,wigner2}. A dusty plasma consists of interacting
microscopic dust particles immersed in an electron-ion plasma. The
dust particles acquire a net charge and the Coulomb interaction
between the dust-particles is shielded by the electron-ion plasma
resulting in a Yukawa or screened Coulomb inter-particle
interaction. The dust particles are confined to a two-dimensional
layer through a combination of gravitational and electrical
forces. By microstructuring a channel in the bottom electrode of
the discharge it is possible to laterally confine the dust
particles as was realized in Refs.
\cite{1homann,2misawa,3liu,4nelser,5sheridan}, the strength of the
1D confinement potential can be varied by the width of the channel
or the potential on the bottom electrode. When the width of the
channel is microstructured into an oscillating function along the
channel it will result in a periodic potential along the channel.

Alternatively, one can confine charged colloids, that move in a
liquid environment containing counterions, into microchannels as
e.g. recently realized experimentally in \cite{6}. In this case
the inter-colloid interaction can be modeled by a screened Coulomb
interaction and the confinement potential is a hard wall
potential. By changing the depth profile of the micro-channel it
has been shown in \cite{7} that the confinement potential can be
tuned into a harmonic potential. Micro-structuring the width of
the channel into an oscillating function along the channel will
result into an additional periodic potential along the channel.

In a previous work \cite{gio04} the ordered configurations of
Yukawa particles confined to Q1D were studied. A phase diagram was
obtained as function of the particle density and inverse Debye
screening length which is a measure of the strength of the
inter-particle interaction. The competition between the lateral
confinement and the screened Coulomb interaction resulted in
different phases where the particles are ordered in chains. The
most well studied phases are the one- and two-chain configurations
where the transition between those two phases occurs through a
zig-zag transition. The latter is a continuous transition as found
theoretically for mono- \cite{gio10} and bi-disperse
\cite{wpferreira,wand10} systems, and experimentally
\cite{4nelser,5sheridan} with a power law dependence on the width
\cite{Candido,gio10}. Here we are interested to investigate how
the phase diagram will be modified when an additional 1D periodic
potential is present. For example, how the zig-zag transition will
be modified by the periodic potential.

The present paper is organized as follows. In Sec. \ref{sec:model}
we describe the model system and methods used in the calculation
of the properties. In Sec. \ref{sec:structure} we present the
results for the different ground state configurations. In Sec.
\ref{sec:phonon} the normal mode spectra for the one and two-chain
regimes are presented for different intensities of the periodic
potential. Our conclusions are given in Sec.
\ref{sec:conclusions}.

\section{THE MODEL }
\label{sec:model}
\begin{figure}[!ht]
\includegraphics[scale = 0.8]{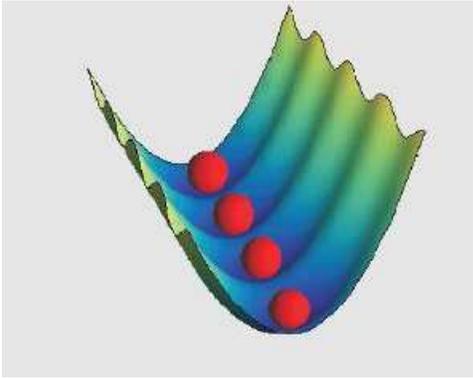}
\caption{(Color online) A sketch of the model system. }\label{fig:sketch}
\end{figure}

Our system consists of identical point-like particles interacting
through a screened Coulomb potential. The particles are allowed to move
in a two-dimensional (2D) plane and are subject to an external
parabolic confinement in the y-direction and a periodic substrate
potential along the $x$-direction. A sketch of the present model
system is shown in Fig. \ref{fig:sketch}. The total interaction energy
of the system is given by
\begin{equation}\label{eq:enegy1}
{\rm{H '= }}\frac{{{\rm{\textit{q}}}^{\rm{2}}
}}{{\varepsilon}}\sum\limits_{i < j} {\frac{{e^{ - | {\vec {r'_i }
- \vec {r'_j } } | /\lambda  } }}{{\left| {\vec {r'_i } - \vec
{r'_j } } \right|}}}  + \sum\limits_i {\frac{1}{2}} m\omega _0^2
{y'_i}^2 + V_0' \sum\limits_i {\cos (\frac{{2\pi x'_i }}{L})}
\end{equation}
\noindent where $\epsilon$ is the dielectric constant of the
medium in which the particles are moving in, $\lambda$ is the
Debye screening length, $V_0'$ is the strength of the periodic
substrate potential, $L$ is the periodicity of the substrate
potential, and $\mathbf{r'}_i=(x'_i,y'_i)$ is the position of the
$i^{th}$ particle. In order to keep in Eq. (\ref{eq:enegy1}) only
the parameters which rule the physics of the system, it is
convenient to write the energy and the distances in units of
$E_{0}=(m \omega_{0}^{2}q^{4}/2\epsilon ^{2})^{1/3}$ and
$r_{0}=(2q^{2}/m \epsilon \omega _{0}^{2})^{1/3}$, respectively,
and the screening parameter $\kappa = r_{0}/\lambda$, where. We
also define the dimensionless strength of the substrate potential
$V_0=V_0'/E_0$ and $\vec{r}=\vec{r'}/r_0$. In so doing, the
expression for the energy is reduced to
\begin{equation}\label{eq:enegy2}
{\rm{H = }} \sum\limits_{i < j} {\frac{{e^{ - \kappa | {\vec {r_i
} - \vec {r_j } } |  } }}{{\left| {\vec {r_i } - \vec {r_j } }
\right|}}}  + \sum\limits_i {y_i}^{2}  + V_0 \sum\limits_i {\cos
(\frac{{2\pi x_i }}{L})}
\end{equation}
As can be observed from Eq. (\ref{eq:enegy2}) the system is a
function of the parameters, $\kappa$, $V_0$, $L$, and the density.
In our numerical calculations $\kappa=1$, which is a typical value
for dusty plasma and colloidal systems. We introduce a distance
$a_0$, which is defined as the distance between particles when
$V_0=0$. The density (\textit{n}) is the ratio between the number
of chains $N_{ch}$ and $a_0$, i.e. $n=N_{ch}/a_0$. In this case,
the system self-organizes in a multichain-like structure
\cite{gio04}.

The model studied in this work is related to the Frenkel-Kontorova
(FK) model, which is a simple one-dimensional model that describes
the dynamics of a chain of particles interacting with nearest
neighbors in the presence of an external periodic potential. This
model was initially introduced in the 1930s by Frenkel and
Kontorova and was subsequently reinvented independently by others,
notably Frank and Van der Merwe \cite{pcm}. It provides a simple
and realistic description of commensurate-incommensurate
transitions when thermal fluctuations are unimportant.

In the present work the quasi-1D character of our system makes it
different from the 1D FK-model. The particles have the additional
freedom to move perpendicular to the chain which leads to a rich
set of new phases. The presence of two length scales in the
FK-model, i.e. the inter-particle distance and the periodicity of
the 1D potential, is the reason of the complex behavior of the
model. The inter-particle potential favors a uniform separation
between the particles, whereas the V(x) tends to pin the particles
at the minima of the periodic potential. This competition between
both interactions is often called frustration or length scales
competition.

The minimum energy configurations are obtained by numerical and
analytical calculations. In the numerical simulations, we
typically considered 100-200 particles, together with periodic
boundary conditions in the unconfined direction in order to mimic
an infinite system. We do not consider  friction in the present
paper. In spite of the primary importance of friction to the
motion of the particles in real systems, the ground state
configurations are not affected by it.

Notice that the substrate is defined in terms of the parameter
$L$. Comparing $L$ and $a_0$, we define here an initially
commensurate (IC) when ($L/a_0=p/q$, with $p$ and $q$ integers)
and initially non-commensurate (INC) regime of the ordered
structures when the ratio $L/a_0$ is a irrational number. It
should be emphasized that in these cases the inter-particle
distance $a_0$ is defined in the absence of a substrate ($V_0=0$).
In the case $V_0\ne0$, it is expected that the mean distance
between particles along a given chain $a$ changes as a function of
$V_0$, driving the system to new commensurate or non-commensurate
configurations.

\section{Ground State Configurations}
\label{sec:structure}

In this section, we present the results obtained analytically and
numerically  for the ground state configurations. In the former,
we calculate the energy per particle for different configurations
as a function of the strength and periodicity of the substrate. We
minimize such expressions with respect to the different distances
between particles. The configuration with lowest energy is the
ground state. In order to predict which structures should be taken
into account in the analytical approach, we also use molecular
dynamic simulations as a complementary tool. The numerical method
can give us some hints about which structures to consider. It
should be noticed that one of the draw backs of the numerical
technique is that in some cases there exist a larger number of
meta-stable states, mainly in the limit of high densities where
the system is found in a multi-chain structure. On the other hand,
the numerical approach is the only way to obtain the ground state
configurations in some incommensurate regimes, which will be
analyzed in the next sections.

We show here that depending on the periodicity of the substrate we
can tune the ground state configuration, induce structural phase
transitions and control the number of chains. This is interesting
from an experimental point of view, since the number of chains can
be associated with the porosity of the system, making it a
controllable filter.

The main features of the present model system can be already seen
in the more simple situations of the single- and two-chain
regimes. For this reason, we limit ourselves to these cases,
because it simplifies the physical interpretation of our results.

\subsection{Single-chain regime}
As an example, we study in this section systems with densities
$n=0.5$ and $n=0.7$, which are found in the single-chain regime
\cite{gio04} in the absence of the substrate $V_0=0$. For $n=0.5$
we consider the commensurate ratios $L/a_0=1$ and $L/a_0=2$, while
for $n=\sqrt{2}/{2}$ we consider the non-commensurate regime with
$L/a_0=\sqrt{2}$.

The simplest IC configuration is the trivial single-chain regime
where each particle is sitted in a minimum of the periodic
potential with $n=0.5$ and $L/a_0=1$. In this case, the
configuration remains the same for any value of $V_0$. Notice that
the cases in which $L/a_0=1/I$, where $I\geq1$ is an integer, will
exhibit the same behavior, since each particle is positioned
exactly in a minimum of the substrate potential. On the other
hand, the case $L/a_0=I$ is very different and the particle
configuration depends strongly on $V_0$, as will be shown in the
next paragraph.

\begin{figure}[h!]
\includegraphics[scale = 0.95]{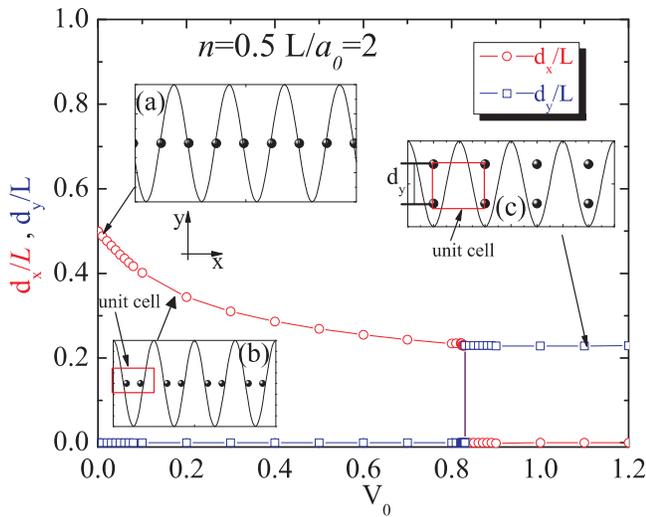}
\caption{(Color online) Nearest neighbor separation between
particles in the $x$- ($d_x$) and $y$-direction ($d_y$) as a
function of $V_0$ for the case $n=0.5$ and $L/a_0=2$. Three
possible configurations are shown as insets.
}\label{fig:n05la2_dist}
\end{figure}

In the IC case with $n=0.5$ and $L/a_0=2$, for small values of
$V_0$ the particles are alternately located at the minimum and at
the maximum of the substrate potential [see inset (a) in Fig.
\ref{fig:n05la2_dist}]. A sufficient increase of $V_0$ forces the
system to a new single-chain configuration in which a pair of
particles is located at each minimum of the substrate [see inset
(b) in Fig. \ref{fig:n05la2_dist}]. A further increase of $V_0$
pushes each pair of particles closer to each other, increasing the
repulsive energy between them.

For a critical value of $V_0$ ($\approx 0.8$) a structural
transition to the two-chain configuration is induced [see inset
(c) of Fig. \ref{fig:n05la2_dist}]. The system changes from a one
to two-chains configuration. In the two-chain configuration the
separation $d_x$ between particles in the x-direction of each
minimum of the substrate is zero, which means particles are
aligned along the y-direction. The separation $d_y$ between chains
does not change as function of $V_0$. In this particular
configuration $d_y$ is ruled only by the competition between the
repulsive interaction between particles and the parabolic
confinement, being independent of the strength of the periodic
potential. The type of transition observed here is different from
the one found in Ref. \cite{gio10}, where the authors demonstrated
that in the absence of a periodic potential and in the presence of
a parabolic confinement we have only continuous transitions from
one to two chains. In our case the one- to two-chains transition
is clearly a first-order phase transition as function of $V_0$.
Notice that in the two-chain regime the system is re-organized in
a final commensurate structure with a new ratio $L/a=1$. We can
define here a commensurate-commensurate transition between
different orders of commensurability.

The expression for the energy per particle which is able to
describe all phases observed in the case with $n=0.5$ and
$L/a_0=2$ and in the case $n=1.0$ and $L/a_0=1$ is given by
\begin{equation}\label{eq:la2n05}
\begin{split}
E = &\frac{n}{2}\sum_j \frac{e^{-2\kappa j/n}}{j}
+ \frac{n}{4}\sum_j \frac{e^{-\frac{2\kappa}{n}\sqrt{[(j-1)+c_x]^2  + c_y^2 }}}
{\sqrt {[(j-1)+c_x]^2  + c_y^2 } }
\\&+ \frac{n}{4}\sum_j \frac{e^{-\frac{2\kappa}{n}\sqrt{(j - c_x)^2  + c_y^2 }}}
{\sqrt {(j-c_x)^2  + c_y^2 } }  + 4 \Big(\frac{c_y}{n}\Big)^2 -
\cos(\pi c_x)
\end{split}
\end{equation}
\noindent where $c_x = d_x /L$ and $c_y = d_y /L$ are,
respectively, the dimensionless separations between particles
within a minimum of the periodic potential along the chain and
perpendicular to it.
\begin{figure}[h!]
\includegraphics[scale = 0.85]{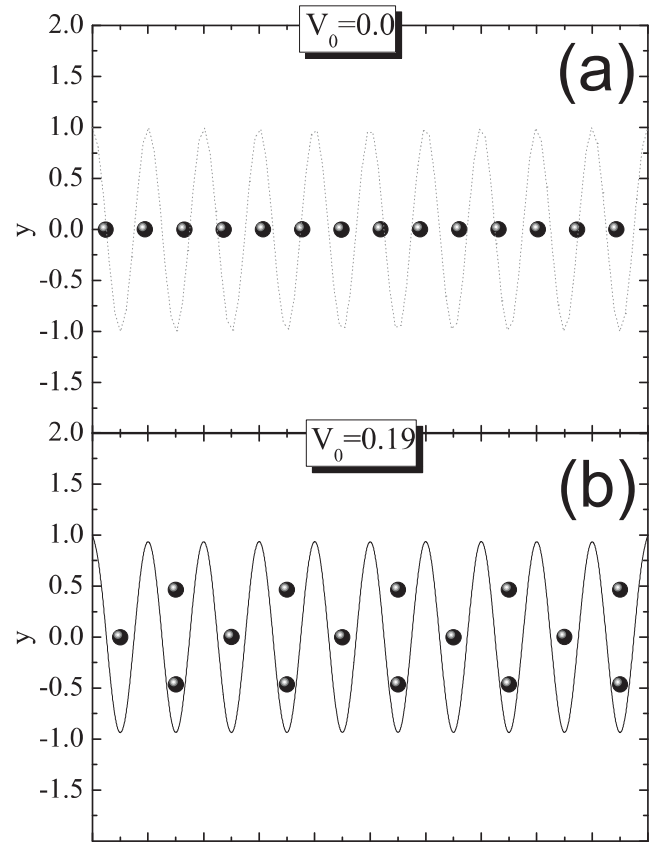}
\caption{ Ground state configurations for the case $n=\sqrt{2}$
and $L/a_0=\sqrt{2}$ for: (a) $V_0=0.17$ (b) $V_0=0.19.$
}\label{fig:n07la15}\end{figure}

Now we discuss the INC regime with $n=\sqrt{2}$ and
$L/a_0=\sqrt{2}$. The same general behavior of previous cases can
be observed here, with several structural transitions ruled by the
strength of the periodic substrate $V_0$ (Fig. \ref{fig:n07la15}).
For a large enough $V_0$ the system can be found in a final
commensurate regime with $L/a\approx1/2$, but now in the
three-chain configuration with particles almost uniformly
distributed over chains.

\subsection{Two-chain regime}
\begin{figure}[h!]
\includegraphics[scale = 0.7]{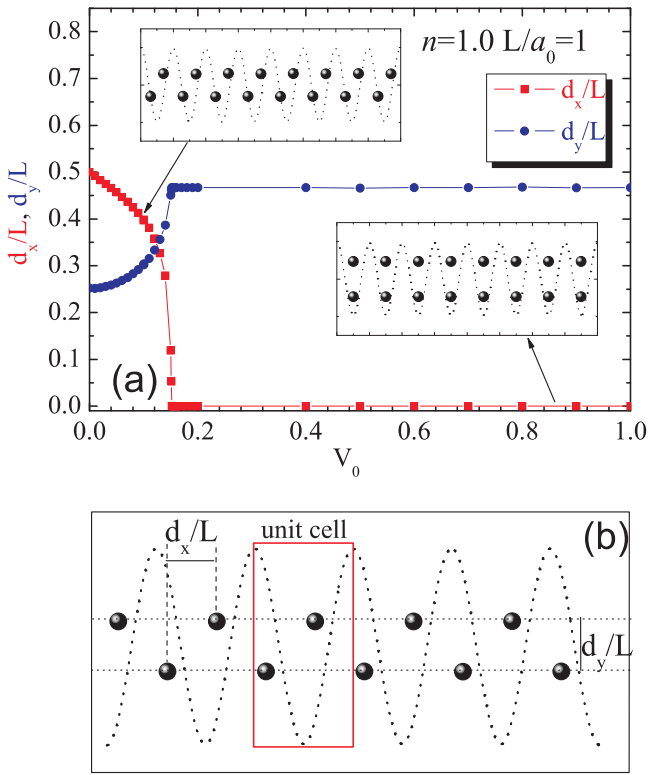}
\caption{ (Color online) (a) Inter-particle separation as function
of $V_0$ for $n=1.0$ and $L/a_0=1.0$. (b) A sketch of the
two-chain configuration with the distances $d_x$ and $d_y$
indicated.} \label{fig:n1la1}\end{figure}

In this section we consider the system with $n=1.0$, where a
two-chain configuration is found as the ground state for $V_0=0$.
Differently from what was observed in the one-chain configuration
($n=0.5$), when $L/a_0=1.0$, the two chain configuration remains,
but the internal structure depends on $V_0$. This is shown in Fig.
\ref{fig:n1la1}(a), where the relevant internal distances [Fig.
\ref{fig:n1la1}(b)] for the arrangement are presented as function
of $V_0$.

For $V_0=0.16$ the system changes from a staggered ($d_{x}\neq0$)
to an aligned ($d_{x}=0$) two-chain configuration, through a
second order (continuous) structural transition, characterized by
a discontinuity in the second derivative of the energy with
respect to $V_0$. Notice that in the case $L/a_0=1.0$, there are
always two particles per period of the substrate potential, and in
such a commensurate phase the system is always found in the
two-chain regime.

\begin{figure}[h!]
\includegraphics[scale = 0.8]{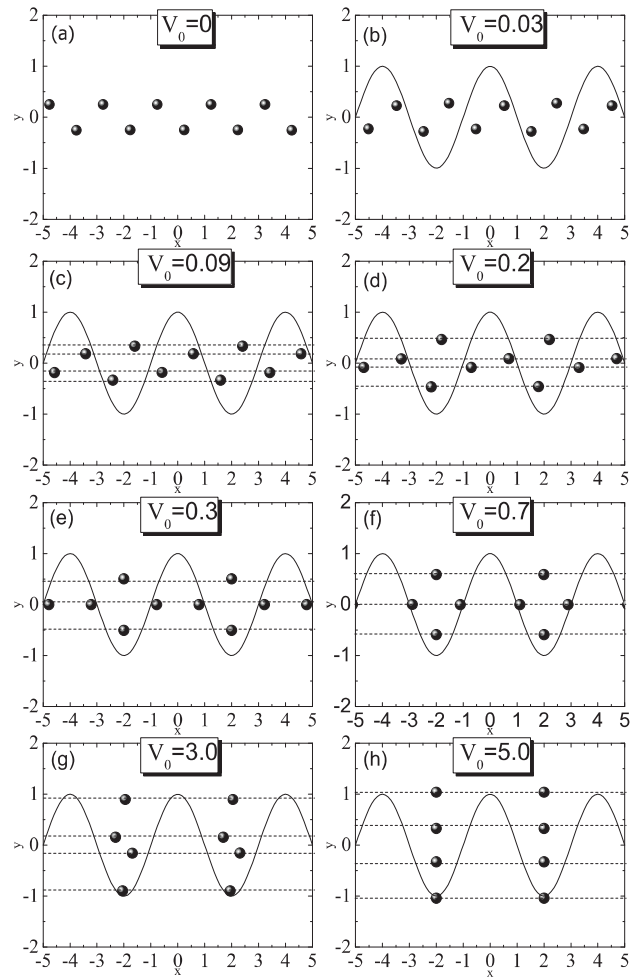}
\caption{\small Ground state configurations for the case $n=1.0$
and $L/a_0=2$ for different values of
$V_0$.}\label{fig:n1la2}\end{figure}

Next, we consider the more interesting case with $n=1.0$ and
$L/a_0=2$. When $V_0$ is increased some unusual configurations
appear as shown in Fig. \ref{fig:n1la2}. Initially the particles
move in the x-direction towards the minima of the periodic
potential and at the same time each chain starts to break up into
two chains [Figs. \ref{fig:n1la2}(c)]. The transition found here
is second-order.

With further increase of $V_0$ the two inner chains move towards
each other [see Fig. \ref{fig:n1la2}(d)] and merge into a single
chain in the center [see Figs. \ref{fig:n1la2} (d,e,f)]. The
particles in the outer chains move towards the minimum of the
periodic potential [see Figs. \ref{fig:n1la2} (d,e,f)]. With
further increase of $V_0$ the pair of particles in the middle
chain are pushed closer to each other and finally form a row of
four particles directed along the y-direction and positioned in
the minimum of the periodic potential [see Figs. \ref{fig:n1la2}
(g,h)]. The configurations presented in Fig. \ref{fig:n1la2}
indicates a tunable porosity of the system as function of $V_0$.
This is very convenient feature if the system is settled to be
used as a filter or sieve, as pointed out in Refs.
\cite{dna1,dna2}, where superparamagnetic colloidal particles were
self-assembled in chain-like structures and used for separation of
DNA molecules.

\begin{figure}[h!]
\includegraphics[scale = 0.55]{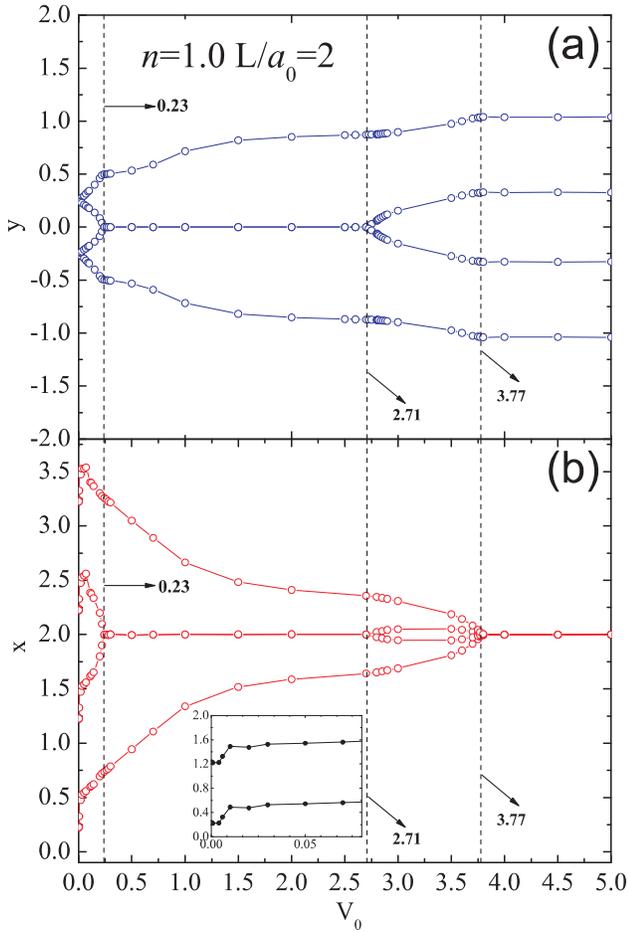}
\caption{\small (Color Online) (a) The lateral $y$-position of
chains for the case $n=1.0$ and $L/a_0=2$. The vertical dotted
lines represents the values of $V_0$ where structural transitions
occur. (b) The particle position in the $x$-direction as function
of $V_0$.}\label{fig:nila2dist}
\end{figure}

The movement of the different particles in the x- and y-direction
as function of $V_0$ is summarized in Fig. \ref{fig:nila2dist},
where structural transitions are indicated by vertical dashed
lines. Three second order structural transitions are observed as
function of $V_0$ and the number of chains varies from
$2\rightarrow4\rightarrow3\rightarrow4\rightarrow4$.

\begin{figure}[h!]
\includegraphics[scale = 0.8]{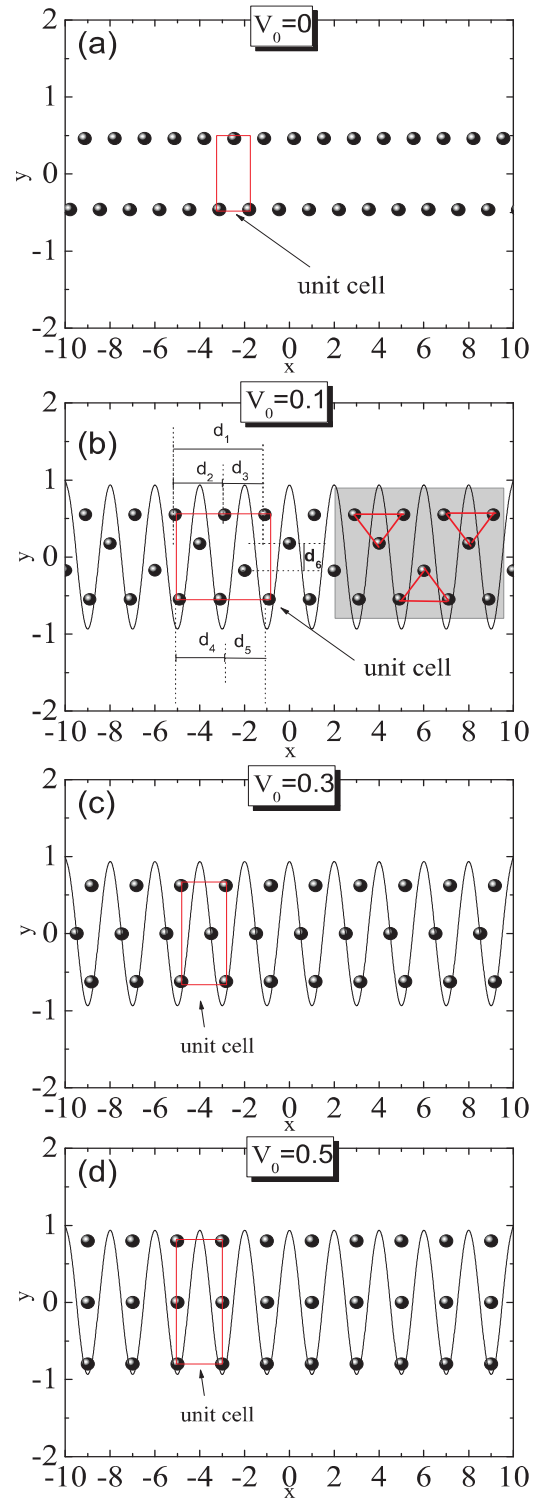}
\caption{(Color online) Ground state configurations for different
values of $V_0$ for $n=1.5$ and $L/a_0=1.5$. In (b) the relevant
distances for the analytical calculation of the energy are
presented. }\label{fig:n15la15}\end{figure}

Finally, we study the case with $n=1.5$ and $L/a_0=1.5$. Here the
commensurate ratio changes according to $V_0$. Initially, for
$V_0=0$, the system is arranged in two chains [see Fig.
\ref{fig:n15la15}(a)], which are displaced with respect to each
other over half the inter-particle distance in each chain. There
are two particles per unit cell, which characterize an initially
commensurate (IC) configuration.

When $V_0$ increases the system transits to a four-chain
configuration through a second or first order structural
transition, with the outer chains having twice as many particles
as the inner chains [Fig. \ref{fig:n15la15}(b)]. Alternatively, we
can also view this configuration as two chains of triangles as
indicated in the shadowed region in Fig. \ref{fig:n15la15}(b). In
this case $d_2> d_3$ and $d_5 > d_4$ and the length of the unit
cell is $d_1=d_2+d_3$. There are six particles in the unit cell,
as in the case $V_0=0$.

\begin{figure}
\includegraphics[scale = 0.9]{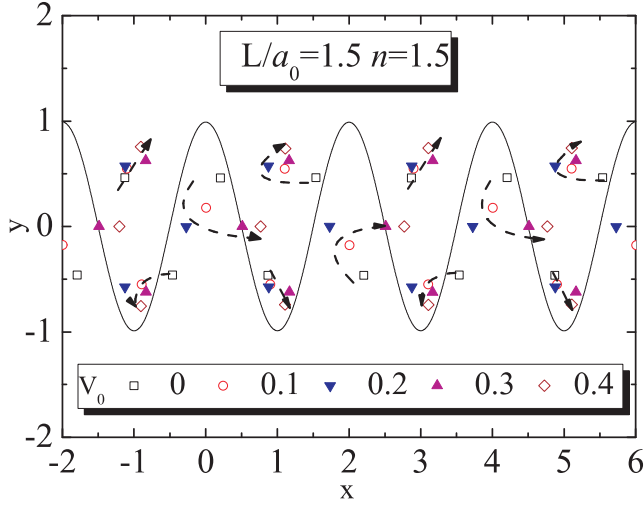}
\caption{\small (Color Online) Particle trajectories for different
values of $V_0$ near the potential minimum for the case $n=1.5$
and $L/a_0=1.5$.}\label{fig:trajectories}
\end{figure}

With further increase of $V_0$, the y-distance $d_6$ between the
internal chains goes to zero and the system changes to the
three-chain configuration [Fig. \ref{fig:n15la15}(c)] with the
same number of particles in each one, and the central chain
shifted by $a/2$ along the x-direction with respect to the outer
chains, which are aligned along the y-direction. Notice that in
this case there are only three particles per unit cell. This is
interesting since the number of normal modes is now half
the one observed for the configuration presented in
Fig. \ref{fig:n15la15}(b), where the number of particles in the unit
cell is six. The reduction of the allowed excitation modes is
controlled by the strength of the periodic potential, and this
can be used as an important feature for possible application in
phononics. For $V_0>0.5$, particles in different chains are all
aligned along the y-direction and located in each minimum of the
periodic substrate [Fig. \ref{fig:n15la15}(d)]. The trajectories
of the different particles in the channel as function of $V_0$ is
visualized in Fig. \ref{fig:trajectories}.

Again, the relation between the periodicity of the substrate and
the distance between particles is different from the case $V_0=0$,
$L/a_0=1$. This is interesting since we can change the
commensurability of the system by changing only the strength of
the substrate potential.

As presented in Figs. \ref{fig:n05la2_dist}(c), \ref{fig:n1la1}
and \ref{fig:n1la2}(h), for a critical value of $V_0$ the present
model system is found in the special configuration where the
particles are aligned along the confinement direction. Such a
$y$-aligned configuration (YAC) occurs if the condition $L/a_0=p$,
where $p$ is an integer ($\geq1$), is satisfied. In this case, if N
is the number of chains of the initial structure ($V_0=0$), then
we find that the number of particles aligned along the
$y$-direction in each minimum of the substrate potential is $N.p$,
which is also the number of chains. The critical value of $V_0$
for which the YAC phase can be induced is obtained by adding the
interaction between particles and confinement energies. A general
expression for the YAC is given by:

\begin{equation}\label{eq:powerlaw1}
\begin{split}
 V = &\frac{n}{{Np}}\sum\limits_j {\frac{{e^{ - kNpj/n} }}{j}  }
 \\& + \frac{{2n}}{{N^2 p^2 }}\sum\limits_{q = 1}^{Np - 1} {\sum\limits_{l = q + 1}^{Np} {\sum\limits_j {\frac{{e^{ - kNp/n\sqrt {j^2  + (p - q)^2 c^2 } } }}{{\sqrt {j^2  + (p - q)^2 c^2 } }}} } }
 \\& + \frac{n}{{N^2 p^2 }}\sum\limits_{q = 1}^{Np - 1} {\sum\limits_{l = q + 1}^{Np} {\frac{{e^{ - k(l - q)cNp/n} }}{{(l - q)c}}} }  \\&+ \frac{{2c^2 Np}}{{n^2 }}\sum\limits_{l = 1}^{Np/2} {(l - 1/2)^2 ,}
\end{split}
\end{equation}
in the case where $N.p$ is even, and

\begin{equation}\label{eq:powerlaw2}
\begin{split}
 V = & \frac{n}{{Np}}\sum\limits_j {\frac{{e^{ - kNpj/n} }}{j}  }
 \\& +\frac{{2n}}{{N^2 p^2 }}\sum\limits_{q = 1}^{Np - 1} {\sum\limits_{l = q + 1}^{Np} {\sum\limits_j {\frac{{e^{ - kNp/n\sqrt {j^2  + (p - q)^2 c^2 } } }}{{\sqrt {j^2  + (p - q)^2 c^2 } }}} } }
 \\& +\frac{n}{{N^2 p^2 }}\sum\limits_{q = 1}^{Np - 1} {\sum\limits_{l = q + 1}^{Np} {\frac{{e^{ - k(l - q)cNp/n} }}{{(l - q)c}}} } \\&+ \frac{{2c^2 Np}}{{n^2 }}\sum\limits_{l = 1}^{Np/2} {l^2 ,}
\end{split}
\end{equation}
if $N.p$ is an odd number.
\begin{figure}
\includegraphics[scale = 0.8]{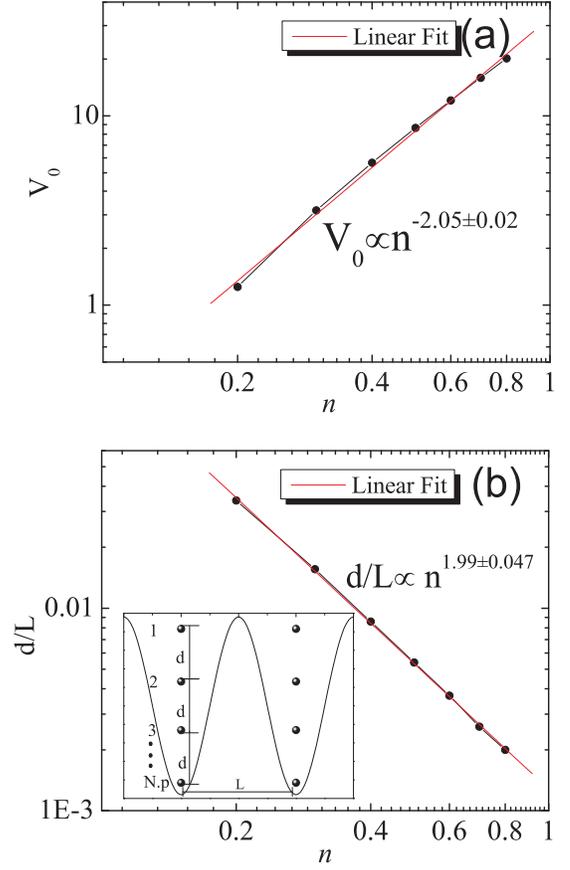}
\caption{(Color online) (a) Critical value of $V_0$ as function of
density for the YAC phase. (b) The distance between the particles
along the y-direction as function of the density in the YAC phase.
The inset shows a sketch of the general groundstate configuration
with all relevant parameters. The red line in both figures is the
linear fit.}\label{fig:powerlaw}
\end{figure}

The critical value of $V_0$ and the separation $d$ between
particles in each minimum are presented in Fig.
\ref{fig:powerlaw}. A sketch of the configuration in each minimum
of the periodic substrate with all relevant parameters is also
shown as insect in Fig.\ref{fig:powerlaw}(b). A power law
dependence of $V_0$ and $d/L$ on the density is found.

\section{Phonon Spectrum}
\label{sec:phonon}

Next, we analyze the $V_0$-dependence of the normal mode spectrum.
We follow the standard harmonic approximation and take into
account the periodicity of the system in the
unconfined direction ($x$-axis).

The present model system is a strictly 2D system where the number
of particle in the unit cell and the number of degrees of freedom
per unit cell determines the number of branches in the phonon
spectrum. If \textit{l} is the number of particles per unit cell,
there will be 2\textit{l} branches in the phonon spectrum, from
which half of those branches correspond to oscillations along the
chain, i.e along the \textit{x} axis we have longitudinal modes,
while the others are associated with vibrations along the
confinement direction (\textit{y} axis transverse modes). If the
particles in the unit cell oscillate in-phase, the mode is
dominantly acoustical, while the opposite out-of-phase oscillation
corresponds to an optical mode. In general, a normal mode can be
classified in one of the following classes: longitudinal optical
(LO), longitudinal acoustical (LA), transverse optical (TO), or
transverse acoustical (TA).

In the harmonic approximation the normal modes are obtained by
solving the system of equations

\begin{equation}
(\omega^2 \delta_{\mu\nu,ij} - D_{\mu\nu,ij})Q_{\nu,j}=0,
\end{equation}

\noindent where $Q_{\nu,j}$ is the displacement of particle $j$ from its equilibrium position in
the $\nu$ direction, $\mu$ and $\nu$ refer to the spatial coordinates $x$ and $y$,
$\delta_{\mu\nu,ij}$ is the unit matrix and $D_{\mu\nu,ij}$ is the dynamical matrix, defined by

\begin{equation}
D_{\mu\nu,ij}=\frac{1}{m}\sum_{u}\phi_{\mu,\nu}(u) e^{-iuqa},
\end{equation}

\noindent where $u$ is an integer assigned to each unit cell. The force constants are given by

\begin{equation}
\phi_{\mu,\nu}(u)=\partial_\mu \partial_\nu \frac{\exp(-\kappa \sqrt{(x-x')^2 +
(y-y')^2})}{\sqrt{(x-x')^2 + (y-y')^2}},
\end{equation}

\noindent with $(x-x')=$ distance between particles along the
x-axis and $(y-y')=$ interchain distance with $(x,y)$ and
$(x',y')$ the equilibrium positions of the particles in the unit
cell, and

\begin{equation}
\phi_{\mu,\nu}(u=0)=-\sum_{u\neq0}\phi_{\mu,\nu}(u).
\end{equation}

\noindent The phonon frequency is given in units of $\omega_0 / \sqrt{2}$. As an example, the
complete dynamical matrix for the one-chain and the two-chain regime are given in Appendix.

The frequencies for the one-chain configuration in the case
$V_0=0$, are given by $\omega_l=\sqrt{A_1}$ for the acoustical
branch and $\omega_t=\sqrt{1+A_2}$ for the optical branch, $A_1$
and $A_2$ are defined in Appendix.

The frequencies for the one-chain configuration when we have $V_0
\neq 0$ and two-chain configuration can be given by:

\begin{equation}
\omega_l  = \sqrt {\frac{1}{4}(B_1  + B_3  \pm \sqrt {B_1 ^2  +
4B_5 B_6  - 2B_1 B_3  + B_3 ^2 }  + sub)}
\end{equation}
for the longitudinal modes, and by

\begin{equation}
\omega_t = \frac{1}{2}\sqrt {4 + B_2  + B_4  \pm \sqrt {B_2 ^2  +
4B_6 B_8  - 2B_2 B_4  + B_4 ^2 } }
\end{equation}
for the transverse modes. The expressions for $B_1,B_2,...,B_6$
and are given in Appendix. Here $sub=8 V_0 \pi^2 cos( \pi c_x)$ is
the term related to the periodic substrate. The wave number $k$
for the one- and the two-chains regimes is in units of $2\pi/L$,
where $L$ is the length of the unit cell in the $x$-direction.

\begin{figure}
\includegraphics[scale = 0.8]{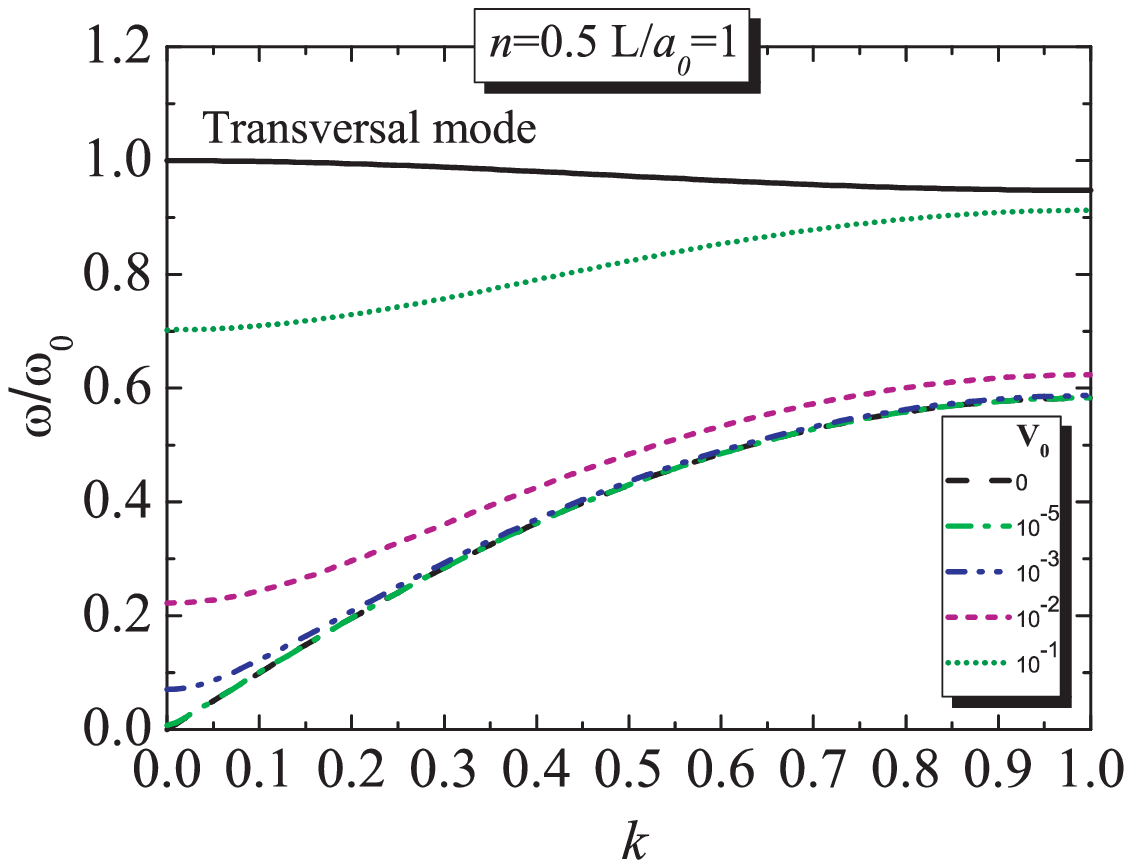}
\caption{(Color online) The phonon spectrum for different values
of $V_0$ in the case $n=0.5$ and $L/a_0=1$.
}\label{fig:modos1}\end{figure}

In Fig. \ref{fig:modos1}(a), the phonon spectrum for the one-chain
configuration is presented for different values of $V_0$, fixed
density $n=0.5$ and $L/a_0=1$. In this case, there is one particle
per unit cell located in each minimum of the substrate resulting
only in one longitudinal mode and one transversal mode. The
frequency of the longitudinal mode increases with increasing
$V_0$, and there is a gap opening at $k=0$. The reason is that the
periodic potential acts locally as a parabolic confinement
potential $V(x) \simeq V_0 \frac{2\pi^2}{L^2} x^2$ with frequency
$\omega=\sqrt{V_0/m} \frac{2\pi}{L}$. The $k=0$ gap corresponds
with this frequency for not too small values of $V_0$. The
transversal mode corresponds to particle oscillations in the
y-direction and is therefore practically independent of $V_0$.
\begin{figure}
\includegraphics[scale=0.8]{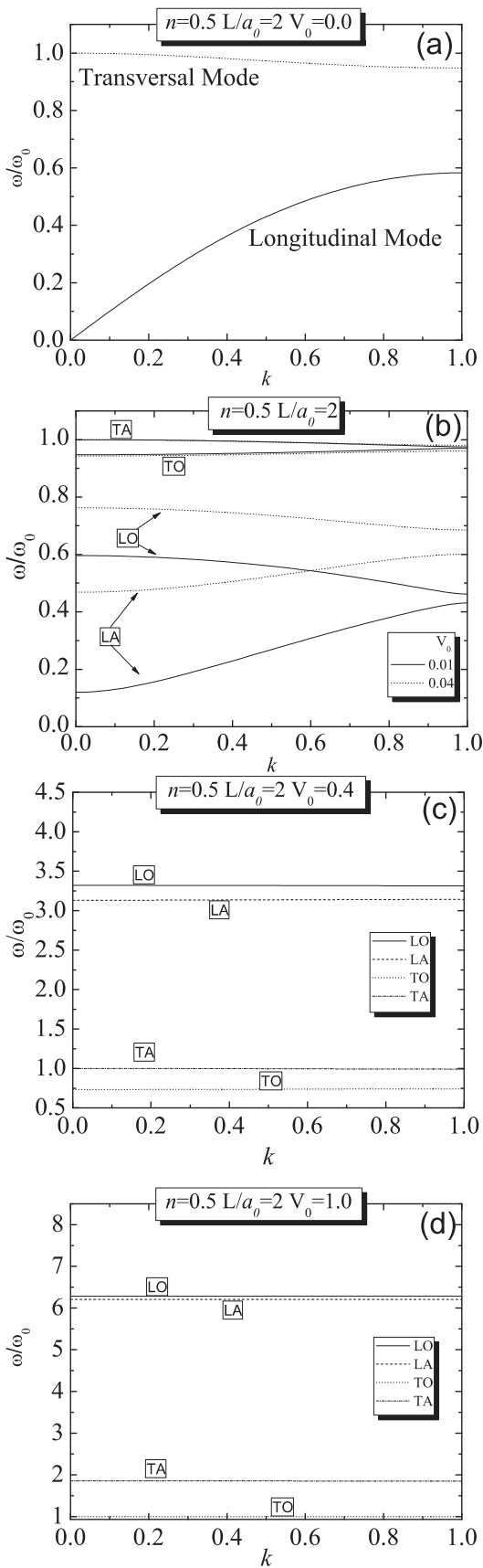}
\caption{The phonon spectrum for different values of $V_0$ in the
case $n=0.5$ and $L/a_0=2$.}\label{fig:modos2}
\end{figure}

In Fig. \ref{fig:modos2}, the dispersion curves for $n=0.5$ and
$L/a_0=2$ are presented for different values of $V_0$. As observed
in Fig. \ref{fig:n05la2_dist}, the presence of the substrate
($V_0\neq0$) modifies the number of particles in the unit cell in
order that the number of branches of the phonon spectrum is
increased as compared to the case $V_0=0$ [Fig.
\ref{fig:modos2}(a)]. For $V_0\neq0$ there are two particles per
unit cell and consequently four branches in the phonon spectrum.
As $V_0$ increases, the frequency of the LA mode also increases,
which can be explained keeping in mind that for low values of
$V_0$, there is a small electrostatic repulsion between neighboring
particles, in order that particles oscillate horizontally without
major difficulties. The opposite behavior is found for the TO
mode, i.e., decreases with increasing $V_0$. The distance between
adjacent particles in the same substrate minimum becomes smaller,
and the repulsive force between them increases and acts as a
retarding force.

The LO mode has a rather different behavior as compared to the TO
mode, i.e., there is a hardening of its frequency when $V_0$
increases, which is a consequence of the larger repulsion due to
the closer proximity between particles. For a sufficiently strong
$V_0$ [ Fig. \ref{fig:modos2}(c)] the normal mode spectrum becomes
discrete, i.e. frequencies become independent on \textit{k}, which
means the group velocity is zero and the modes become localized.

As commented previously, for $V_0\geq0.8$ there is a structural
phase transition to the two-chain configuration with particles
aligned along the y-direction (YAC phase) in each minima of the
substrate [Fig. \ref{fig:n05la2_dist}]. Again, due to the strong
confinement the modes are localized, and a discrete spectrum is
found [Fig. \ref{fig:modos2}(c)] with large frequencies for the
longitudinal modes. The transversal modes also present an increase
in the frequency due the large confinement in the y-direction.

\begin{figure}
\begin{center}
\includegraphics[scale=0.8]{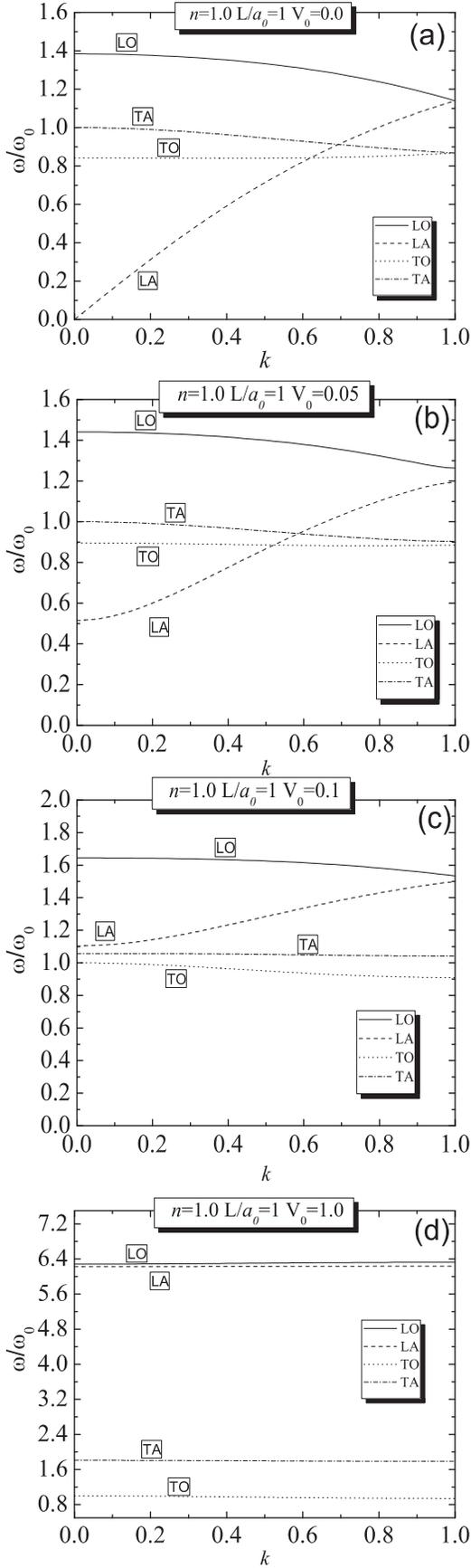}
\caption{The phonon spectrum for different values of $V_0$ in the
case $n=1$ and $L/a_0=1$. }\label{fig:modos4}
\end{center}
\end{figure}

Now we discuss the dispersion curves for the system with density
$n=1.0$ and $L/a_0=1$, [Fig. \ref{fig:modos4}]. As presented
earlier [Fig. \ref{fig:n1la1}], particles remain in the two-chains
configuration for all $V_0$ with changes only in the internal
structure. Again the substrate potential induces gaps in the
normal mode frequencies as presented in Fig. \ref{fig:modos4}. The
TA, TO, LA and LO modes increase with increasing $V_0$.

In the case of the LA mode, for low values of $V_0$ particles are
not aligned, having more freedom to oscillate in the horizontal
direction. When $V_0$ increases, the electrostatic force becomes
larger (particles are now aligned) making oscillations along the
channel harder.

The LO mode also increases with increasing $V_0$. This is a
consequence of the strength of the substrate potential, which trap
particles in their equilibrium positions, reducing the out phase
oscillations of the particles. The TO frequency increases
slightly, since the out-of-phase motion is more difficult to
occur. The TA frequency branch is almost independent of $V_0$
because it corresponds to oscillations in the $y$-direction and is
therefore determined by the harmonic confinement potential with
frequency $\omega_0$.

For the YAC phase, $V_0 > 0.16$, the normal mode spectrum becomes
discrete [Fig. \ref{fig:modos4}(d)], as in the case $n=0.5$,
$L/a=2.0$. The modes are almost constant due to the strong
confinement potential imposed by the substrate and the harmonic
trap.

\section{Conclusions}
\label{sec:conclusions}
We investigated the structural and dynamical properties of a
two-dimensional system of repulsive particles confined by a
parabolic channel and submitted to a one-dimensional periodic
potential (substrate). The ground state configurations were
obtained analytically and numerically, where for the latter we
used molecular dynamics simulations. The phonon spectrum were
also calculated analytically for the one- and two-chain
configurations through the harmonic approximation.

The main features of the structure and normal mode spectrum were
studied (for different densities) as a function of the periodicity
($L$) and strength ($V_0$) of the substrate, which are
experimentally tunable parameters in systems like e.g. colloids
in the presence of a periodic light field composed of two
interfering laser beams. An interesting set of ground state
configurations with controllable porosity is observed mainly as a
function of $V_0$, through several first or second order
structural transitions. The structures are mainly ruled by the
fact that particles tend to go to the minima of the periodic
substrate, modifying the symmetry of the ordered structures.
However, for small $V_0$ the inter particle repulsive
interaction dominates and the particles are found over all possible
positions in the periodic potential, including regions near to the
maxima. For large $V_0$, particles are more and more attracted to
the wells of the periodic potential.

For some specific cases we found structural transitions where the
number of particles in the unit cell of the periodic system is
changed, implying e.g. a different number of branches in the phonon
spectrum, which is an interesting aspect of the dynamical behavior
of the system, specially for applications in phononics.

The normal mode frequencies depend on the linear density of the
system, periodicity and strength of the periodic substrate. We
observed gaps in the phonon spectrum, which indicate that there
are frequencies blocked by the crystal. For $V_0$ beyond a critical
value and for specific values of the ratio $L/a_0$
the system is found in a special configuration were particles are
aligned in each minimum of the periodic substrate and
perpendicular to the $x$-direction. For such a
configuration the normal mode frequencies become independent of
the wave vector and the modes localize into a small frequency of interval.

\begin{acknowledgments}
JCNC, WPF, GAF and FMP were supported by the Brazilian National
Research Councils: CNPq and CAPES and the Ministry of Planning
(FINEP). FMP was also supported by the Flemish Science Foundation
(FWO-Vl).
\end{acknowledgments}

\appendix*
\section{}
\label{app:appendixA}
The matrix $\omega^2 \textbf{I} -
\textbf{D}$ (where $\textbf{I}$ is the unit matrix and
$\textbf{D}$ is the dynamical matrix) is used in the calculation
of the normal modes for the one- and two-chains configurations.
The dynamical matrix for one chain configuration when $V_0=0$ is:
$$
\left[
\begin{array}
{rrrr}
\omega^{2}-A_{1}&  0\\
0&   (\omega^{2}-\omega_0^{2})-A_{2}\\
\end{array}\right]
\label{matrix} ,
$$

\noindent where the quantities $A_1$ and $A_2$ are given by:
\begin{equation}
\begin{split}
A_1&=\sum_{j=1}^{\infty}n^{3}\frac{e^{-\kappa j/n}}{j^{3}}\Big[2+\frac{2\kappa j}{n}+\frac{\kappa^2 {j}^{2}}{n^{2}}\Big][1-cos(k\pi{j})] \\
&+ {V_{0}}{\pi n}^2cos(\pi j)
\end{split}
\end{equation}

\begin{equation}
A_2=\sum_{j=1}^{\infty}n^{3}\frac{e^{-\kappa j/n}}{j^{3}}\Big[1+\frac{\kappa {j}}{n}\Big][1-cos(k\pi{j})]
\end{equation}

The dimensionless wave number $k$ is in units of $2\pi/L$. The
dynamical matrix to one-chain $V_0\neq0$ and two-chains
configuration is:

$$
\left[\begin{array}
{rrrr}
\omega^{2}-B_{1} - sub&  0& -B_{5}&0\\
0&   \Delta\omega^{2}-B_{2} &0&-B_{6}\\
-B_{7}&0&\omega^{2}-B_{3} - sub&0\\
0&-B_{8}&0&     \Delta\omega^{2}-B_{4}\\
\end{array}\right]\label{matrix} ,
$$

\noindent where $\Delta\omega^{2} = \omega^2 -\omega_{0}^{2}$. The quantities $B_1$, $B_2$, $B_3$,
$B_4$, $B_5$ and $B_6$ are given by:

\begin{equation}
\begin{split}
B_1&=\sum_{j=1}^{\infty}n^{3}\frac{e^{-2\kappa
r/n}}{(2r)^{3}}\Big[(j-c_x)^2\Big(\frac{3}{r^2}+\frac{6\kappa
}{nr}+\frac{4\kappa^2}{n^{2}}\Big)\\&-(1+\frac{2\kappa
r}{n})\Big]+\sum_{j=1}^{\infty}n^{3}\frac{e^{-2\kappa j
/n}}{(2j)^{3}}\Big[2+\frac{4\kappa
j}{n}+\frac{(2\kappa j)^{2}}{n^{2}}\Big]\\
&\times[1-exp(ikjL)] \\
\end{split}
\end{equation}

\begin{equation}
\begin{split}
B_2&=\sum_{j=1}^{\infty}n^{3}\frac{e^{-2\kappa
r/n}}{(2r)^{3}}\Big[\frac{3c_y^2}{r^2}+\frac{4 \kappa^2 c_y^2}{n^2}+\frac{6\kappa
c_y^{2}}{nr}-(1+\frac{\kappa r}{n})\Big]\\
&- \sum_{j=1}^{\infty}n^{3}\frac{e^{-2\kappa
j/n}}{(2j)^{3}}\Big[1+\frac{2\kappa j}{n}\Big][1-exp(ikjL)],
\end{split}
\end{equation}

\begin{equation}
\begin{split}
B_3&=\sum_{j=1}^{\infty}n^{3}\frac{e^{-2\kappa
r_1/n}}{(2r)^{3}}\Big[(j-1+c_x)^2\Big(\frac{3}{r^2}+\frac{6\kappa
}{nr}+\frac{4\kappa^2}{n^{2}}\Big)\\&-(1+\frac{2 \kappa
r_1}{n})\Big] +\sum_{j=1}^{\infty}n^{3}\frac{e^{-2\kappa j
/n}}{(2j)^{3}}\Big[2+\frac{4\kappa
j}{n}+\frac{(2\kappa j)^{2}}{n^{2}}\Big]\\
&\times[1-exp(ikjL)] \\
\end{split}
\end{equation}

\begin{equation}
\begin{split}
B_4&=\sum_{j=1}^{\infty}n^{3}\frac{e^{-2\kappa
r_1/n}}{(2r_1)^{3}}\Big[\frac{3c_y^2}{r_1^2}+\frac{4 \kappa^2 c_y^2}{n^2}+\frac{6\kappa
c_y^{2}}{nr_1}-(1+\frac{\kappa r_1}{n})\Big]\\
&- \sum_{j=1}^{\infty}n^{3}\frac{e^{-2\kappa
j/n}}{(2j)^{3}}\Big[1+\frac{2\kappa j}{n}\Big][1-exp(ikjL)]
\end{split}
\end{equation}

\begin{equation}
\begin{split}
B_5&=\sum_{j=1}^{\infty}n^{3}\frac{e^{-2\kappa
r/n}}{(2r)^{3}}\Big[(j-c_x)^2\Big(\frac{3}{r^2}+\frac{6\kappa
}{nr}+\frac{4\kappa^2}{n^{2}}\Big)\\&-(1+ \frac{2\kappa
r}{n})\Big][exp(ikL(j-c_x))] \\
\end{split}
\end{equation}

\begin{equation}
\begin{split}
B_6&=\sum_{j=1}^{\infty}n^{3}\frac{e^{-2\kappa
r/n}}{(2r)^{3}}\Big[\frac{3c_y^2}{r^2}+\frac{4 \kappa^2 c_y^2}{n^2}+\frac{6\kappa
c_y^{2}}{nr}-(1+\frac{\kappa r}{n})\Big]\\&\times[exp(ikL(j-c_x))],
\end{split}
\end{equation}

\begin{equation}
\begin{split}
B_7&=\sum_{j=1}^{\infty}n^{3}\frac{e^{-2\kappa
r_1/n}}{(2r)^{3}}\Big[(j-1+c_x)^2\Big(\frac{3}{r^2}+\frac{6\kappa
}{nr}+\frac{4\kappa^2}{n^{2}}\Big)\\&-(1+\frac{2 \kappa
r_1}{n})\Big][exp(ikL(j+c_x))] \\
\end{split}
\end{equation}

\begin{equation}
\begin{split}
B_8&=\sum_{j=1}^{\infty}n^{3}\frac{e^{-2\kappa
r_1/n}}{(2r_1)^{3}}\Big[\frac{3c_y^2}{r_1^2}+\frac{4 \kappa^2 c_y^2}{n^2}+\frac{6\kappa
c_y^{2}}{nr_1}-(1+\frac{\kappa r_1}{n})\Big]\\&\times[exp(ikL(j+c_x))],
\end{split}
\end{equation}

\noindent where $r=\sqrt{(i - c_x)^2 + c_y^2}$, $r_1=\sqrt{(i - 1
+ c_x)^2 + c_y^2}$, the dimensionless wave number $k$ is in units
of $2 \pi/L$, $i=\sqrt{-1}$ and $sub=8 V_0 \pi^2 cos( \pi c_x)$.

\end{document}